\documentclass[12pt,preprint]{aastex}

\shorttitle{Debris Disks around WDs}
\shortauthors{Kilic et al.}

\begin{document}

\title{Debris Disks Around White Dwarfs:  The DAZ Connection}

\author{Mukremin Kilic\altaffilmark{1,2}, Ted von Hippel\altaffilmark{2}, S. K. Leggett\altaffilmark{1,3}, and D. E. Winget\altaffilmark{2}}

\altaffiltext{1}{Visiting Astronomer at the Infrared Telescope Facility, which is operated by the University of Hawaii under Cooperative Agreement no. NCC 5-538 with the National Aeronautics and Space Administration, Office of Space Science, Planetary Astronomy Program.}
\altaffiltext{2}{The University of Texas at Austin, Department of Astronomy, 1 University Station
C1400, Austin TX 78712, USA; kilic@astro.as.utexas.edu}
\altaffiltext{3}{Joint Astronomy Centre, 660 N. A'ohoku Place, University Park, Hilo HI 96720, USA}

\begin{abstract}

We present near-infrared spectroscopic observations of 20 previously known DAZ white dwarfs obtained
at the NASA Infrared Telescope Facility. Two of these white dwarfs (G29-38 and GD362) are known to display
significant K-band excesses due to circumstellar debris disks. Here we report the discovery of excess K-band
radiation from another DAZ white dwarf, WD0408-041 (GD56).
Using spectroscopic observations, we show that the excess radiation cannot be
explained by a stellar or substellar companion, and is likely to be caused by a warm debris disk.
Our observations strengthen the connection between the debris disk phenomena and the observed metal abundances in
cool DAZ white dwarfs. However, we do not find any excess infrared emission from the most metal rich DAZs with $T_{\rm eff}=$
16000 -- 20000 K. This suggests that the metal abundances in warmer DAZ white dwarfs
may require another explanation.
\end{abstract}


\section{Introduction}

The presence of planets around solar type stars suggests that some white dwarfs should have planetary systems as well. 
Even though planets within 5 AU of their parent stars will most likely not survive the red giant phases,
much of the outer planets, Kuiper belt objects, and Oort-like comet clouds are predicted to survive around white dwarf stars (Debes \& Sigurdsson
2002). Near and mid-infrared searches for stellar and substellar companions to white dwarfs have resulted in the
discovery of late type dwarfs and two brown dwarfs (Farihi et al. 2005), but no planetary systems yet (Mullally et al.
2006).
The search for the effects of reflex orbital motion in changing light travel times caused by possible planetary companions
to pulsating white dwarfs (Winget et al. 2003) has provided strong limits on one object so far,
G117-B15A (Kepler et al. 2005). Finding a Jupiter-size planet around an intrinsically faint, Earth-size white dwarf would herald
a new age of direct studies of extra solar planets.

An easier way to detect the signatures of planetary systems around white dwarfs is via debris disks.
Planets in previously stable orbits around a star undergoing mass loss may become unstable, and some of these systems may result
in close encounters which could result in tidal stripping of a parent body that would end up in a circumstellar debris disk
(Debes \& Sigurdsson 2002; Jura 2003). Until recently, there was only a single white dwarf known to have a circumstellar
debris disk, G29-38 (see Chary et al. 1999). G29-38 is a pulsating DAZ white dwarf with a hydrogen rich atmosphere that has trace amounts
of metals. Infrared excess around this white dwarf was initially thought
to be due to a brown dwarf companion (Zuckerman \& Becklin 1987), however, follow-up observations in
the near- and mid-infrared showed that the excess emission is caused by a debris disk. Reach et al. (2005) discovered silicate
emission around G29-38 and could explain the observed mid-infrared spectrum by a combination of amorphous olivine, amorphous carbon,
and crystalline forsterite. These observations suggest a relatively recent disruption of one or perhaps a few comets or
asteroids around this white dwarf.
 
The presence of metals in the photosphere of G29-38 is probably related to the presence of a dust disk around the star.
Possible scenarios for the explanation of the metal abundances in DAZ white dwarfs include accretion from the interstellar
medium (ISM; Dupuis et al. 1992), cometary impacts (Alcock et al. 1986), and accretion of asteroidal material from a surrounding
debris disk (Graham et al. 1990). If accretion from the ISM is responsible for the observed metal abundances, we would expect to
find cool DAZs only near interstellar clouds since the diffusion timescales for the metals in their photospheres
($\sim$years) are shorter than the cloud crossing timescales ($\geq$10$^4$ years).
Yet Zuckerman et al. (2003) did not find any correlation between the locations and relatively large numbers of DAZs
with the locations and amount of ISM present within the local bubble.

The discovery of the most massive (1.24$M_{\odot}$) and metal rich white dwarf currently known, GD362, by Gianninas
et al. (2004) and Kawka \& Vennes (2005; 2006) stimulated a search for a debris disk around it. GD362 is the most extreme DAZ
star with nearly solar iron and magnesium abundances (log(Fe/H) = -4.5 and log(Mg/H) = -4.8). 
A breakthrough came eighteen years after the discovery of a debris disk around G29-38, when Becklin et al. (2005) and Kilic et al.
(2005) discovered significant excess infrared radiation from GD362. The observed near- and mid-infrared excess around this star
could only be explained by a circumstellar debris disk. These recent results support the idea that accretion from a surrounding
debris disk can explain the metal abundances observed in cool DAZ white dwarfs.

Calcium abundances\footnote{[Ca/H] is the logarithm of the Ca/H abundance ratio by numbers.} for known DAZs (Zuckerman et al. 2003;
Gianninas et al. 2004; Berger et al. 2005) range from [Ca/H]= -5.2 to -12.7.
With several stars being more metal rich than G29-38 ([Ca/H]=-6.9), they
provide a unique opportunity to test if the observed metal abundances can be explained with debris disks similar to G29-38 and GD362.
With a goal of determining the fraction of DAZs with detectable near-infrared excess emission from warm debris disks, we obtained
near-infrared spectroscopy of 20 DAZ white dwarfs including G29-38 and GD362.
Our observations are discussed in \S 2, while an analysis of the spectroscopic data and results from this analysis are
discussed in \S 3.

\section{Observations}

\subsection{Photometry}

All of the DAZ white dwarfs identified by Zuckerman et al. (2003) and Koester et al. (2005) are bright enough to be detected in the
Two Micron All Sky Survey (2MASS). However, the 2MASS photometry in the $H$ and $K$ bands is inaccurate for the majority of the
objects, making the search for infrared excess around these objects unreliable.
Figure 1 presents 2MASS photometry versus effective temperatures for single DAZ white dwarfs studied by Berger et al. (2005), along with
the predicted sequences for DA (solid line) and DB white dwarfs (dashed line) from Bergeron et al. (1995).
The 2MASS photometry for G29-38 and the IRTF photometry for GD362 (filled triangles) are also plotted.
Several objects are only detected in the $J$ and $H$ bands, but not in the $K$ band, and are therefore not included in
the bottom panel.
Even though the large error bars in 2MASS photometry prevent any definitive conclusion, it is clear from this figure that
several objects, most notably GD56, show excess $K$ band flux at the $2-3 \sigma$ level. 

\subsection{Spectroscopy}

We used the 0.8--5.4 Micron Medium-Resolution Spectrograph and Imager (SpeX; Rayner et al. 2003) on the 3m NASA Infrared
Telescope Facility (IRTF) and 0.5$\arcsec$ slit to obtain a resolving power of
90--210 (average resolution of 150) over the 0.8--2.5 $\mu$m range.
Our observations were performed under conditions of thin cirrus and partly cloudy skies between November 2 -- 7, 2005.
To remove the dark current and the sky signal from the data, the observations were taken in two different
positions on the slit (A and B) separated by 10$\arcsec$.
The total exposure times for individual objects ranged from 16 minutes to 96 minutes.
Internal calibration lamps (a 0.1W incandescent lamp and an Argon lamp) were used
for flat-fielding and wavelength calibration, respectively. In order to correct for telluric features and flux calibrate the spectra,
nearby bright A0V stars were observed at an airmass similar to the target star observations.
We used an IDL-based package, Spextool version 3.2 (Cushing et al. 2004), to reduce our data (see Kilic et al. 2005 for details).
Using the nearby A0V star observations and the XTELLCOR package (Vacca et al. 2003), we created a telluric spectrum for each
A0V star observation, and used these spectra to flux calibrate and correct the telluric features in the white dwarf spectra.

\section{Results}

Figure 2 presents the flux calibrated spectra of our DAZ sample (black lines; ordered in $T_{\rm eff}$) and their respective DA white dwarf
model atmospheres (red lines; kindly made available to us by D. Koester and D. Saumon). The resolution of the model white dwarf
spectra were not matched to the instrumental resolution in order to show the predicted locations of the Paschen lines (0.955,
1.005, 1.094, 1.282, and 1.876$\mu$m). Changing the resolution of these models would only make the hydrogen lines
shallower but not change the overall shape of the expected continuum.
The observed white dwarf spectra are affected by strong telluric features between $1.35 - 1.45 \mu$m, $1.80 - 2.05
\mu$m, and longward of $2.5 \mu$m (a typical telluric spectrum observed at the IRTF is presented in Figure 3 of Kilic et al. 2005).
Therefore, weak features observed in several stars in these wavelength ranges are likely to be due to telluric correction problems.

We derived synthetic colors of all 20 objects in our sample using our infrared spectra. Since our observations were performed
under non-photometric conditions, the absolute flux level in these spectra cannot be trusted. Nevertheless, the relative flux level
of the spectra, $J-H$ and $H-K$ colors should not be affected by the non-photometric conditions. The photometric error bars were
calculated from the observed scatter of the spectra in individual frames.
Effective temperatures, gravities, calcium abundances, 2MASS photometry, and our synthetic colors
for these objects are presented in Table 1. A comparison of the 2MASS colors with our synthetic colors shows that
they are consistent within the errors.
Figure 3 presents synthetic colors versus temperatures, and $J-H$ versus $H-K$
color for our sample.
The predicted sequences for DA (solid line) and DB white dwarfs (dashed line) from Bergeron et al. (1995) are also shown.
Our IRTF observations improved the photometry for these objects significantly, enabling us to use the spectroscopy and photometry
to identify near-infrared excess.

A comparison of the spectroscopy and photometry for the DAZ white dwarfs in our sample shows that the majority of the stars do not
show any excess emission in the near-infrared. However, the two previously known debris disk white dwarfs (G29-38 and GD362) plus
WD0408-041 (hereafter GD56) show significant excess in the K-band. The observed spectrum of GD56 is consistent with a typical 14000 K white
dwarf in the J band, but it starts to deviate from a DA white dwarf model in the H band, and shows more infrared excess than either of
the previously known debris disk white dwarfs. GD56, G29-38, and GD362 are easily distinguishable from the rest of the DAZs in our sample
in Figure 3.

There are two other objects with possible slight infrared excesses, WD1015+161 and WD1116+026. WD1015+161 observations may have been
affected by a nearby object, which is located 2.1$\arcsec$ away and $\sim$0.2 mag fainter than the
white dwarf at $J$. We obtained a noisy spectrum
of this object that did not reveal any spectral features. The overall flux distribution of this nearby object could be approximated by
an $\sim$8000 K blackbody. Hence the observed slight excess for WD1015+161 ($T_{\rm eff}=19300 K$) in the K band may be caused by
this nearby object. WD1116+026 may show slight excess in the K band as well, though our synthetic photometry shows that the excess in
$H-K$ color is less than 2$\sigma$. This white dwarf appeared to be a single star in our guider and spectral images.
Therefore, WD1116+026 ([Ca/H]=-7.3) may be another warm debris disk case, but the observed excess and errors
are consistent with photospheric flux from the white dwarf.

We note that the observed spectrum of WD0245+541 is better fitted with a 5190 K helium rich white dwarf model atmosphere (dashed line in
the last panel in Figure 2) rather than a DA model atmosphere (solid line in the same panel). Only weak H$\alpha$ is seen in the spectrum
of cool DAs with $T_{\rm eff}\sim5000$ K, and infrared photometry is often necessary to determine the atmospheric composition of cool
white dwarfs (see Bergeron et al. 2001 for a detailed discussion).
Greenstein \& Liebert (1990) obtained an H$\alpha$ equivalent width measurement of 0.3\AA\ for WD0245+541.
Their classification of this star as a DA white dwarf relies only on optical spectroscopy.
Our near-infrared spectrum of this object favors a helium rich atmosphere model,
therefore WD0245+541 is more likely to be a DZA instead of a DAZ.

\section{Discussion}

Figure 4 shows the IRTF spectra of GD56, G29-38, and GD362, along with the expected near-infrared fluxes for each star (dashed lines).
The observed excess in GD56 is very similar to the excess seen in G29-38 and GD362. The expected flux levels from a 14400 K
blackbody (normalized to the observed GD56 spectrum in the J-band) in the H and K$_S$ bands are 0.27 and 0.17 mJy, respectively.
The difference between the observed and expected flux from the star is about 0.05 mJy in the H band, and 0.18 mJy in the K band.
Using log $g=$ 7.8 and $T_{\rm eff}=$ 14400 K (Berger et al. 2005) and Bergeron et al. (1995) models, we estimate the absolute
K-band magnitude of GD56 to be 11.43 mag. If the observed
near-infrared excess is due to a late type dwarf companion, this would correspond to $M_{\rm K}\sim$ 11.38 mag, an early L dwarf
(Leggett et al. 2002). We used an L3 dwarf template from the IRTF Spectral Library (Cushing et al. 2005) plus the normalized
14400 K blackbody to attempt to match the observed excess in the K band (green line, top panel).
Adding an L3 dwarf to a 14400 K blackbody creates spectral features from 1.3 to 2.5 $\mu$m that are prominent,
yet not seen in the spectrum of GD56, and hence a cool dwarf companion cannot explain the excess seen
between 1.5 -- 2.5 $\mu$m.
Kilic et al. (2005) could fit the observed K band excess around GD362 with a 700 K blackbody.
Likewise, Reach et al. (2005) found a best-fit blackbody temperature of 890 K for the debris disk around G29-38.
The top panel in Figure 3 shows that the excess around GD56 can be explained with an $\sim$890 K blackbody as well.
Therefore, the best explanation for the H and K band excess in GD56 is a circumstellar dust disk heated by the white dwarf.

GD56 was included in Farihi et al.'s (2005) search for substellar companions around white dwarfs. They obtained
$K$-band photometry for $\sim$1/3 of their sample and used 2MASS photometry for the rest of the objects.
Due to low signal-to-noise 2MASS K-band photometry, they did not report a near-infrared excess around GD56 (J. Farihi 2006, private
communication). Their search for excess around 371 white dwarfs resulted in the discovery of late type stars and a brown dwarf,
but no debris disks. Our IRTF search for near-infrared excess around cool DAZs has revealed two new debris disks.
Our sample was restricted to 20 stars (50\% of all known single DAZs) due to the positional constraints imposed by the time of the
observations and the declination limit of the telescope. We observed all known DAZs with $0<\alpha<11$ and $19<\alpha<24$ hours and
$\delta>-17^o$. 

A $Spitzer/IRAC$ search for mid-infrared excess around 4 DAZs by von Hippel et al.
(2006), 4 more DAZs by Debes et al. (2006, private communication), and an additional 17 DAZs by Farihi et al. (2006,
private communication) resulted in the discovery of two more debris disks around cool
DAZs ($T_{\rm eff}<10000$ K), increasing the number of debris disks around DAZ white dwarfs to 5.
There is some overlap between our IRTF program and the Spitzer programs; the total number of DAZs observed so far in near- or
mid-infrared is 35. We estimate that the fraction of cool DAZs with detectable debris disks is 14\% (5 out of 35 stars).
Only 3 of the debris disks are warm enough to be detected in the K-band, and hence the fraction of K-band detectable
debris disks is $\sim$9\%. 
These discoveries suggest a growing connection between cool DAZ white dwarfs and circumstellar debris disks.

\subsection{Possible Trends}

Figure 5 shows the calcium abundances in cool DAZ white dwarfs as a function of effective temperature from Berger et al. (2005).
Objects with IRTF near-infrared spectroscopy (filled circles), and Spitzer 4.5 and 8$\mu$m photometry (filled triangles; von Hippel
et al. 2006) are also shown. The rest of the objects from the Berger et al. (2005) sample are shown as stars. Objects with
circumstellar debris disks are marked with open circles.

None of the six objects with $T_{\rm eff}>$16000 K that were observed at the IRTF and with Spitzer/IRAC show any excess emission
from warm debris disks. Even though we cannot rule out the existence of cool debris disks around these objects, their high metal
abundances require continuous, fairly high accretion rates from surrounding media (Koester et al. 2005; Koester \& Wilken 2006).
The diffusion timescales in a typical DAZ with 0.6$M_{\odot}$ and
$T_{\rm eff}\geq15000$ K are shorter than $\sim$3 days (Paquette et al. 1986). Photospheric metals in these stars have to be
replenished on a daily basis. If the metals were accreted from circumstellar debris disks, we should have been able to detect such
disks around these objects unless all six of them have optically thin, smaller, or edge on disks. Five of these warm DAZs are in
fact more metal rich than GD56 and G29-38.
Radiative levitation can explain the observed metal abundances in hotter white dwarfs, though it is predicted to be
insignificant below 20000 K. 

All three white dwarfs with $T_{\rm eff}\sim 10000 - 15000$ and [Ca/H] $\geq-7.1$ have detectable debris disks around them. In addition, 
WD1116+026 ($T_{\rm eff}=12200 K$ and [Ca/H]$=-7.3$) has a slight (questionable) K-band excess that can be explained by a cooler,
or more distant debris disk. Its calcium abundance is 2.5 times lower than that of G29-38, therefore unlike the warm DAZ stars
mentioned above, a cooler/distant debris disk can actually explain the lower metal abundance of WD1116+026.
There is only one more white dwarf satisfying the above $T_{\rm eff}$ and [Ca/H] criteria
in Berger et al. (2005) sample. WD1150-153 ($T_{\rm eff}=12800 K$
and [Ca/H]$=-6.7$) has more calcium than GD56 and G29-38, and therefore, it is likely to have a debris disk around it.
HE0106-3253 ($T_{\rm eff}=15700 K$ and [Ca/H]$=-6.4$) has surface temperature and calcium abundance similar to warm DAZ white dwarfs
without detectable debris disks. Spectroscopic or photometric observations of both of these objects are required
to search for excess flux in the infrared. 

We did not detect any near-infrared excesses from debris disks in our K band observations of DAZs with $T_{\rm eff}<14000$ and
[Ca/H] $<-7.5$. Using Spitzer/IRAC observations, von Hippel et al. (2006) found excess 4.5 and 8$\mu$m emission around a white
dwarf with $T_{\rm eff}=9700 K$ and [Ca/H] $=-$7.6. 
This object is included in Figure 1 and has $J-H=0.11\pm0.06$ mag and $H-K=-0.03\pm0.08$ mag as measured by 2MASS.
Its 2MASS colors are consistent with the predicted colors for a 9700 K white dwarf.
Von Hippel et al. (2006) also observed a cooler DAZ white dwarf with $T_{\rm eff}=8619 K$ and two orders of magnitude smaller
calcium abundance and did not find any excess up to 8 $\mu$m. 
The lack of infrared excess around this latter object does not mean that it does not have a debris disk, as it may have a
colder, more distant debris disk that would explain the lower calcium abundance and would show up at longer wavelengths.
Four out of 10 DAZs with $T_{\rm eff}$ = 9000 -- 15000 K and [Ca/H] $>$ -8 have debris disks detectable with the current
precision of our observations.

\section{Conclusions}

Our near-infrared spectroscopic observations of 20 cool DAZs resulted in the discovery of one more white dwarf with a
circumstellar debris disk. The observed $H$ and $K$-band excess around GD56 is similar to the infrared excess seen around
the previously known DAZs with circumstellar debris disks (G29-38 and GD362). Due to its $T_{\rm eff}$ and calcium abundance
(i.e. its position in Figure 5), we postulate that another DAZ white dwarf, WD1150-153, is likely to have a $K$-band detectable
circumstellar debris disk. 

The short lifetime of the hot dust (700--900 K) observed around several DAZ white dwarfs may suggest that the hot dust
comes from populations of colder, longer-lived reservoirs of comets or other debris that
are more massive than the hot dust clouds. Debris disks around main sequence stars often have excesses at 24$\mu$m, but hardly
ever at 8$\mu$m (e.g. Chen et al. 2005). DAZ white dwarfs may well behave the same way. The cool reservoir model (a cloud
of dust at $\sim$ 120 K) may be an example of the kind of dust cloud that the recent IRAC surveys could not detect around cool DAZs,
but a sensitive longer wavelength survey could.
The fraction of known single DAZs with near- or mid-infrared (up to 8$\mu$m) excesses is 14\%.
Mid-infrared photometry of more DAZs at longer wavelengths will likely bring this fraction up. Accretion from these debris
disks would also explain the photospheric metal abundances observed in DAZ white dwarfs.

Our understanding of debris disks around white dwarfs will benefit greatly from mid-infrared spectroscopy.
$Spitzer$ spectroscopy of G29-38 revealed a silicate feature around this object suggesting a cometary or asteroidal origin for
the debris disk (Reach et al. 2005). Mid-infrared spectroscopy of GD56 along with the other debris disk white dwarfs will be
needed to check if all of them show silicate emission at 10 $\mu$m.
These observations will help our understanding of the formation and evolution of debris disks around white dwarfs.

\acknowledgements
We would like to thank our referee, Jay Farihi, for helpful suggestions that greatly improved the article.
This material is based upon work supported by the National Science Foundation under grant AST-0307315 and the NASA grant
NAG5-13094. This publication makes use of data products from the Two Micron All Sky Survey, which is a joint project of the
University of Massachusetts and the Infrared Processing and Analysis Center/California Institute of Technology, funded by
the National Aeronautics and Space Administration and the National Science Foundation.

\clearpage
\begin{deluxetable}{lrcrcrrrrr}
\tabletypesize{\scriptsize}
\tablecolumns{9}
\tablewidth{0pt}
\tablecaption{Near-Infrared Photometry of Cool DAZs}
\tablehead{
\colhead{Object}&
\colhead{$T_{\rm eff}$(K)}&
\colhead{log $g$}&
\colhead{[Ca/H]}&
\colhead{$J_{2MASS}$}&
\colhead{$J-H_{2MASS}$}&
\colhead{$H-K_{2MASS}$}&
\colhead{$J-H_{IRTF}$}&
\colhead{$H-K_{IRTF}$}\\
 & & & & $(mag)$ & $(mag)$ & $(mag)$ & $(mag)$ & $(mag)$
}
\startdata
WD0032$-$175& 9235& 8.0 & $-$10.2& 14.79 $\pm$ 0.04 &0.03 $\pm$ 0.08 & 0.03 $\pm$ 0.12 & 0.04 $\pm$ 0.04 & $-$0.02  $\pm$ 0.04\\
HS0047+1903& 16600 & 7.8 & $-$6.1 & 16.33 $\pm$ 0.11 & \nodata & \nodata & $-$0.10 $\pm$ 0.09 & $-$0.10 $\pm$ 0.15\\
WD0208+396& 7201& 7.9 & $-$8.8 & 13.83 $\pm$ 0.02 & 0.16 $\pm$ 0.04 & 0.08 $\pm$ 0.05 & 0.17 $\pm$ 0.03 & 0.01 $\pm$ 0.03\\
WD0235+064& 11420 & 7.9 & $-$9.0 & 15.69 $\pm$ 0.07 & $-$0.22 $\pm$ 0.21 & \nodata & $-$0.05 $\pm$ 0.04 & $-$0.04 $\pm$ 0.08\\
WD0243$-$026& 6798 & 8.2 & $-$9.9 & 14.68 $\pm$ 0.04 & 0.09 $\pm$ 0.06 & 0.11 $\pm$ 0.10 & 0.21 $\pm$ 0.03 & 0.01 $\pm$ 0.04\\
WD0245+541 & 5190 & 8.2 & $-$12.7 & 13.87 $\pm$ 0.02 & 0.33 $\pm$ 0.05 & 0.08 $\pm$ 0.06  & 0.25 $\pm$ 0.03 & 0.07 $\pm$ 0.04\\
HS0307+0746& 10200 & 8.1 & $-$7.6 & 16.39 $\pm$ 0.14 & 0.36 $\pm$ 0.27  & \nodata &  0.08 $\pm$ 0.02 & $-$0.02 $\pm$ 0.02\\
WD0408$-$041 & 14400 & 7.8 & $-$7.1 & 15.87 $\pm$ 0.06 & $-$0.12 $\pm$ 0.14 & 0.55 $\pm$ 0.22 & 0.16 $\pm$ 0.04 & 0.56 $\pm$ 0.04\\
G29-38 & 11600& 8.1 & $-$6.9 & 13.13 $\pm$ 0.03 & 0.06 $\pm$ 0.04 & 0.39 $\pm$ 0.04 &\nodata & \nodata \\
GD362 & 9740 & 9.1 &$-$5.2 &16.16 $\pm$ 0.09&\nodata &\nodata & 0.04 $\pm$ 0.05 & 0.21 $\pm$ 0.04 \\
WD0543+579 & 8142 & 8.0 & $-$10.3 & 15.52 $\pm$ 0.07 & 0.08 $\pm$ 0.15 & 0.11 $\pm$ 0.22  & 0.02 $\pm$ 0.08 & $-$0.05 $\pm$ 0.12\\
WD0846+346 & 7373 & 8.0 & $-$9.4 & 15.89 $\pm$ 0.07 & 0.28 $\pm$ 0.13 & 0.18 $\pm$ 0.21 & 0.31 $\pm$ 0.04 & 0.03 $\pm$ 0.04\\
WD1015+161 & 19300 & 7.9 & $-$6.3 & 16.13 $\pm$ 0.09 &0.01 $\pm$ 0.24 & 0.12 $\pm$ 0.31 & 0.02 $\pm$ 0.03 & 0.08 $\pm$ 0.03\\
WD1116+026 & 12200 & 7.9 & $-$7.3 & 14.75 $\pm$ 0.04 & 0.02 $\pm$ 0.06 & 0.12 $\pm$ 0.12  &  $-$0.02 $\pm$ 0.05 & 0.06 $\pm$ 0.07\\
WD1858+393 & 9470 & 8.0 & $-$7.8 & 15.53 $\pm$ 0.05 & 0.09 $\pm$ 0.10 & 0.19 $\pm$ 0.17 & 0.09 $\pm$ 0.03 & $-$0.03 $\pm$ 0.05\\
HS2132+0941 & 13200 & 7.7 & $-$7.7 & 16.09 $\pm$ 0.08 &0.12 $\pm$ 0.21 & \nodata & 0.05 $\pm$ 0.06 & $-$0.05 $\pm$ 0.09\\
WD2149+021 & 17300 & 7.9 & $-$7.6 & 13.20 $\pm$ 0.02 &$-$0.08 $\pm$ 0.04 & $-$0.11 $\pm$ 0.05 & $-$0.05 $\pm$ 0.03 & $-$0.06 $\pm$ 0.03\\
HE2221$-$1630& 10100 & 8.2 & $-$7.6 & 15.80 $\pm$ 0.06 & 0.02 $\pm$ 0.16 & $-$0.06 $\pm$ 0.14 &  0.10 $\pm$ 0.05 & 0.07 $\pm$ 0.04\\
HS2229+2335 & 18600 & 7.9 & $-$6.3 & 16.16 $\pm$ 0.09 &\nodata &\nodata & 0.02 $\pm$ 0.07 & 0.00 $\pm$ 0.08\\
HE2230$-$1230 & 20300 & 7.7 & $-$6.3 & 16.36 $\pm$ 0.11 & $-$0.08 $\pm$ 0.26 & \nodata &  $-$0.09 $\pm$ 0.08 & $-$0.07 $\pm$ 0.12\\
\enddata
\tablecomments{$T_{\rm eff}$, log $g$, and [Ca/H] are from Berger et al. (2005) and Gianninas et al. (2004).}
\end{deluxetable}

\clearpage
\begin{figure}
\plotone{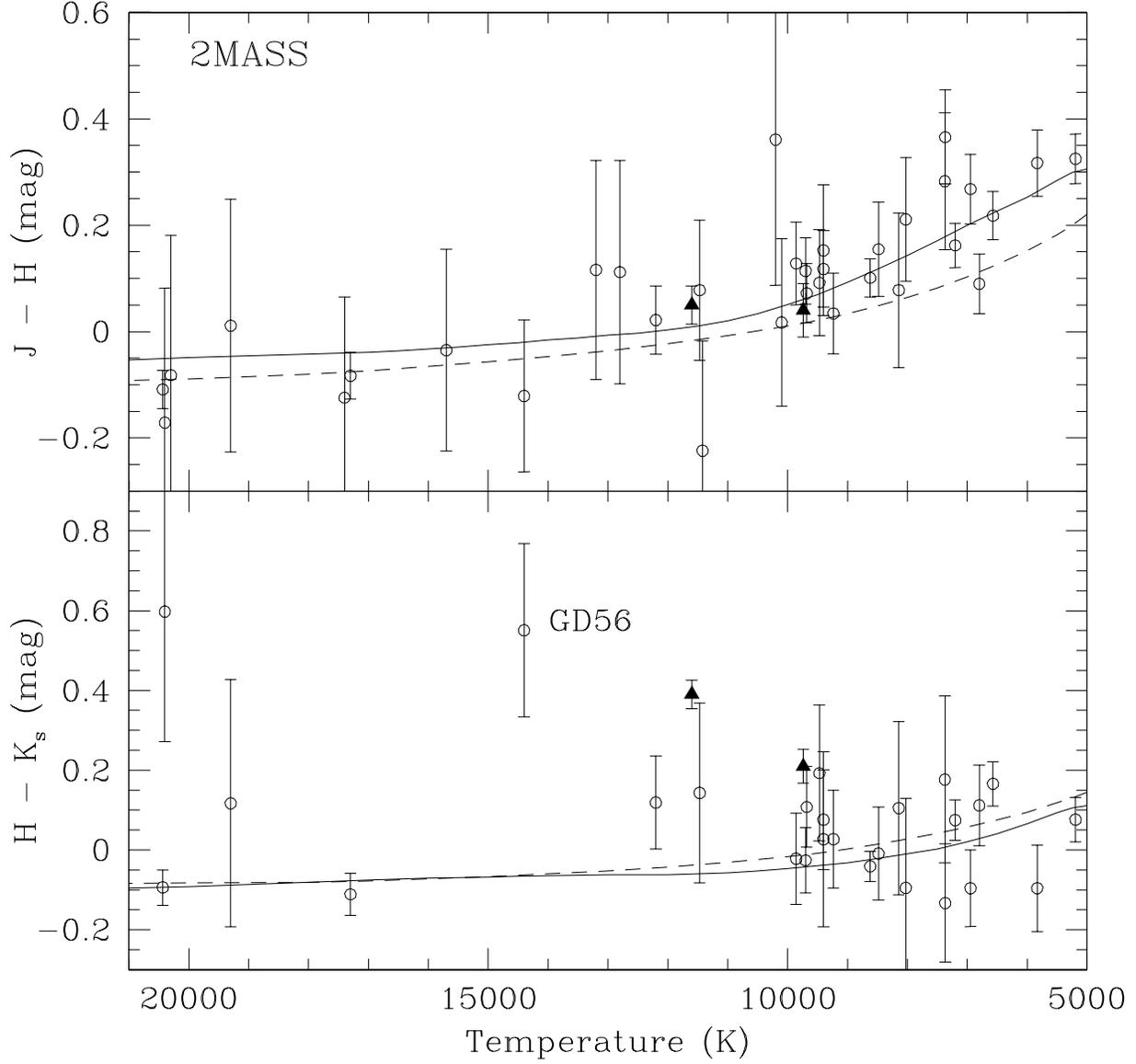}
\caption{$J-H$ and $H-K$ 2MASS colors versus temperature for single DAZ white dwarfs studied by Berger et al. (2005;
open circles). The predicted sequences for DA (solid line) and DB (dashed line) white dwarfs and the colors for G29-38 and
GD362 (filled triangles) are also shown. GD56, a likely debris disk candidate from 2MASS photometry, is labeled.}
\end{figure}

\clearpage
\begin{figure}
\epsscale{1.5}
\plottwo{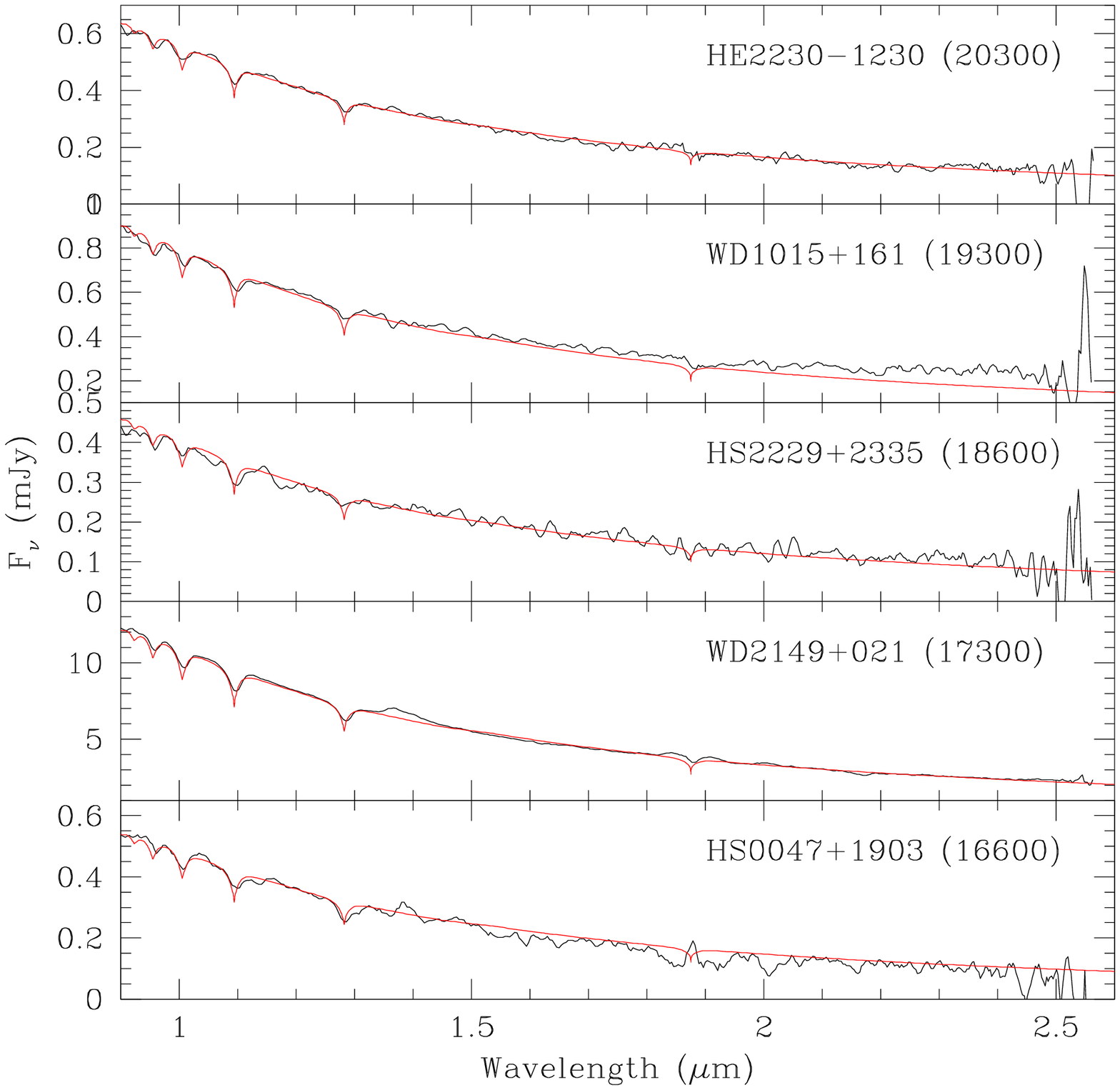}{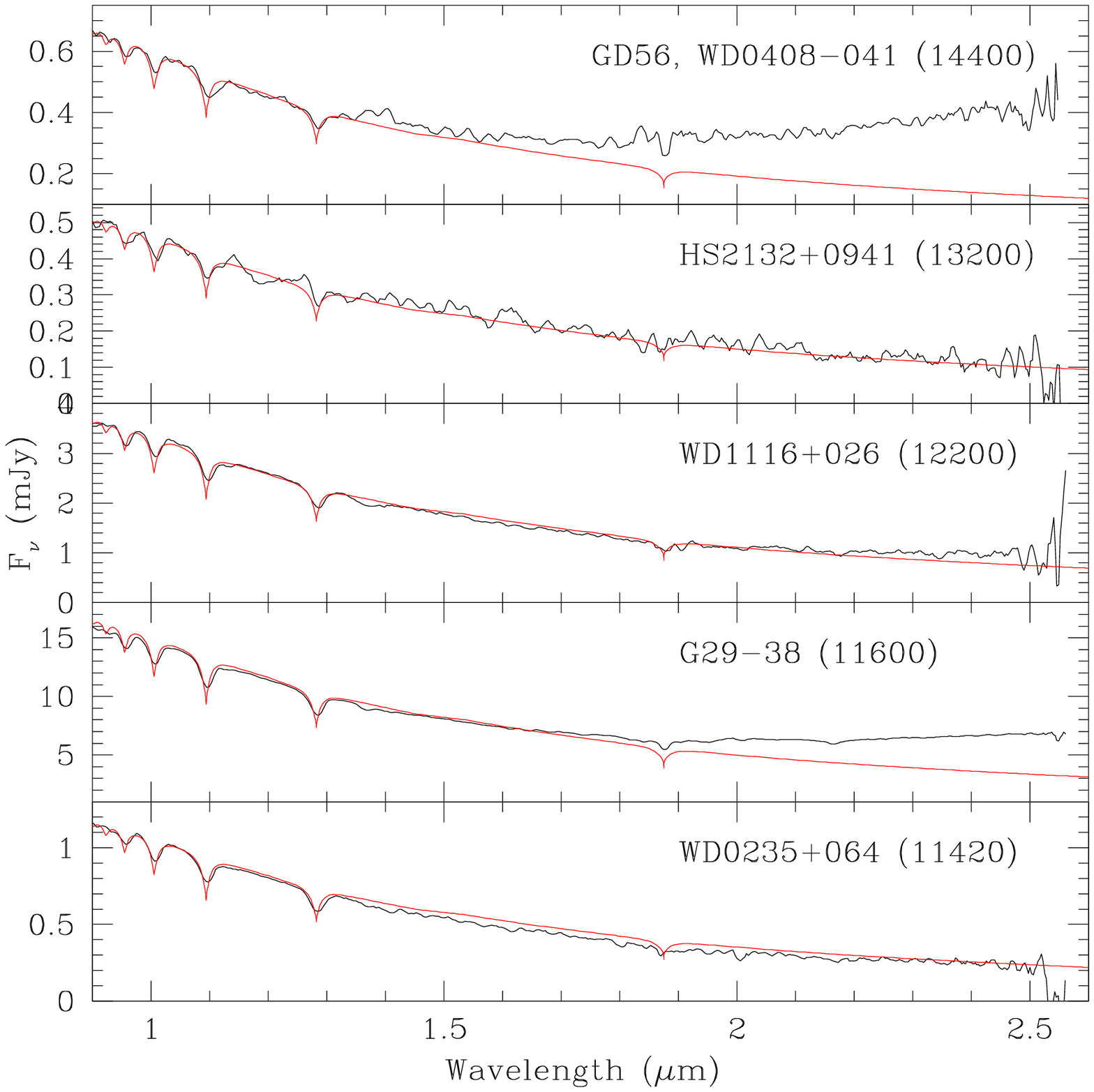}
\caption{Flux calibrated spectra of the DAZ white dwarfs in our sample (black lines; ordered in $T_{\rm eff}$) compared to
models (red lines). }
\end{figure}

\clearpage
\begin{figure}
\figurenum{2}
\epsscale{1.5}
\plottwo{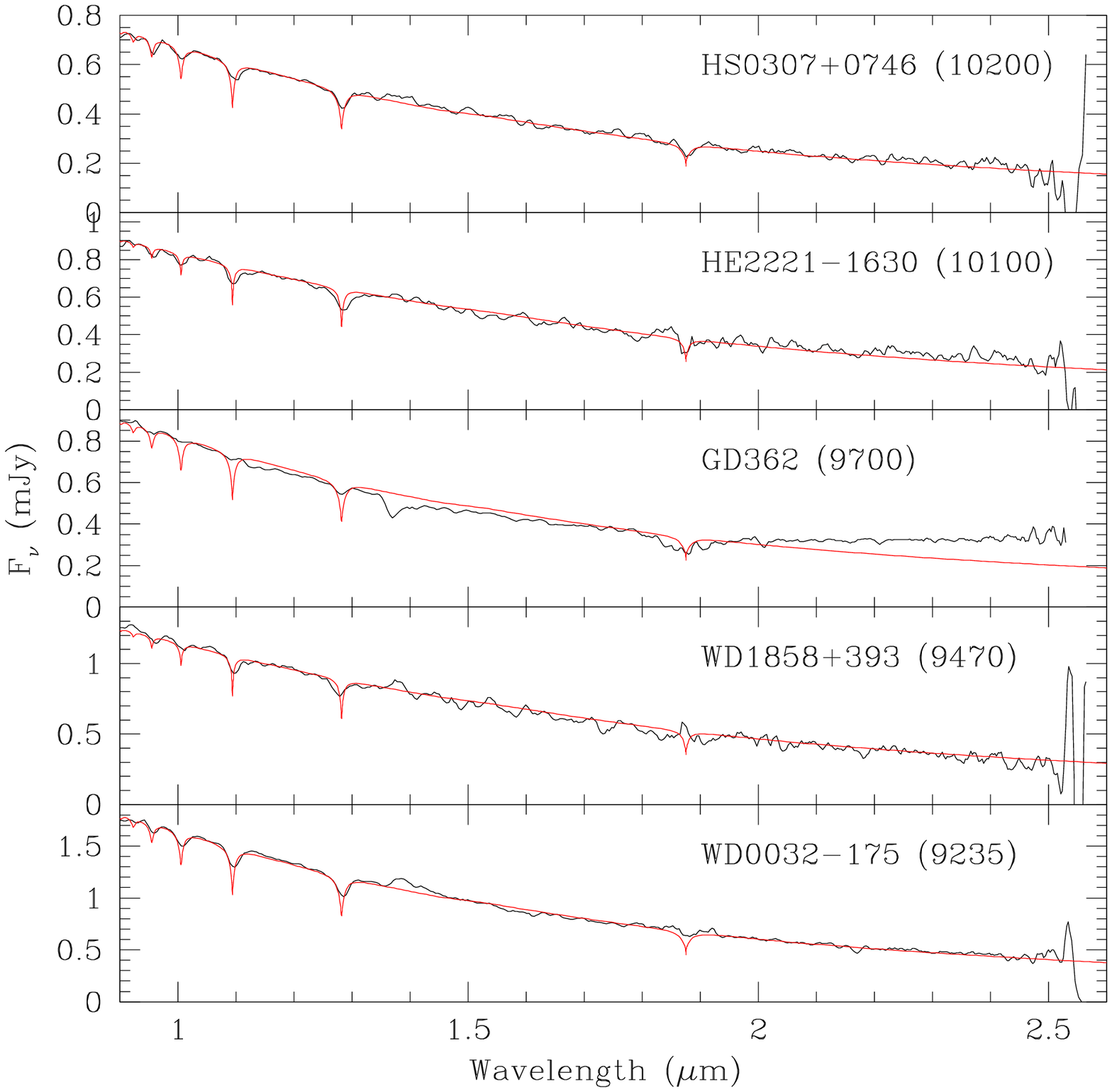}{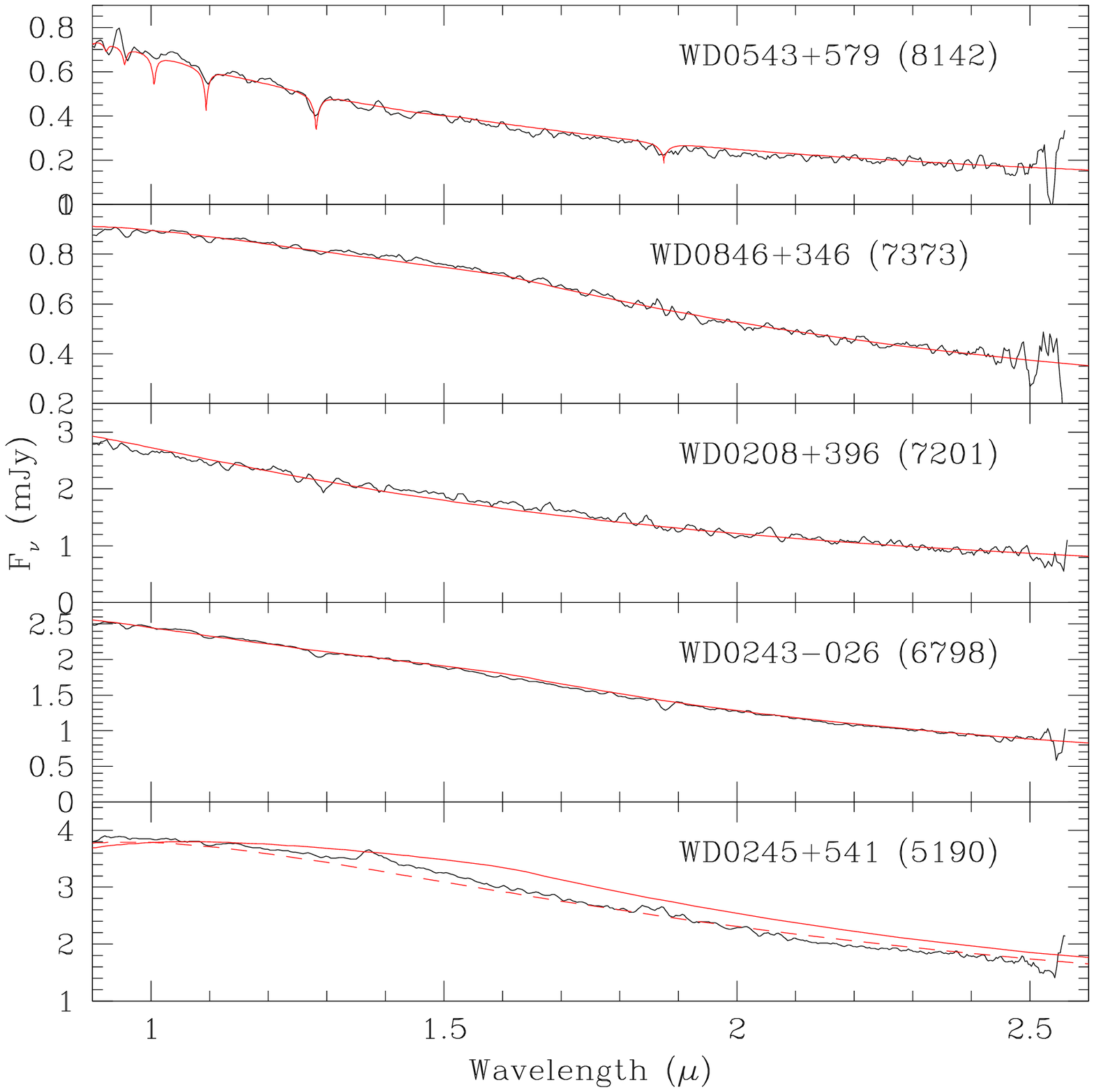}
\caption{cont.}
\end{figure}

\clearpage
\begin{figure}
\epsscale{1.25}
\plottwo{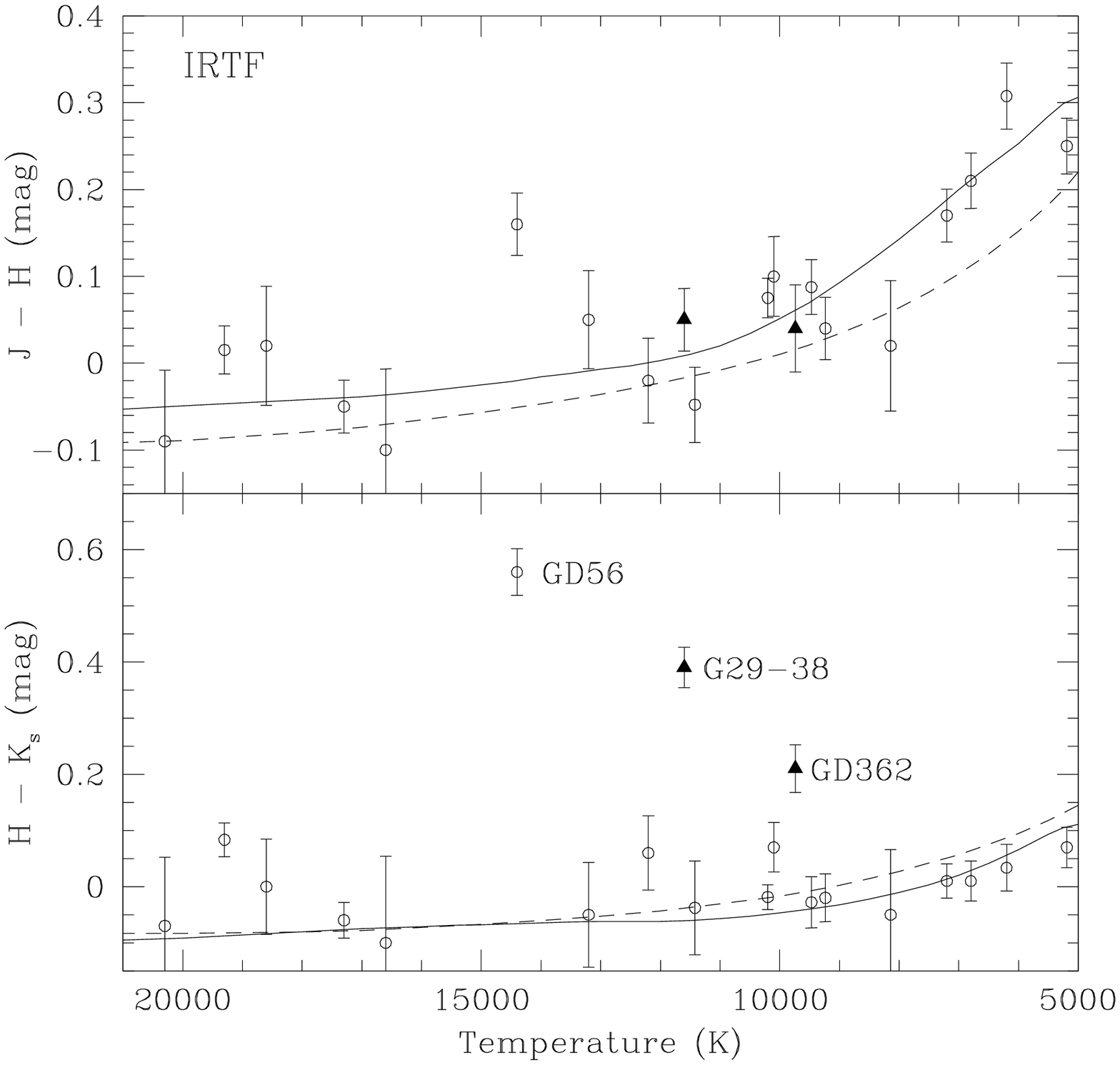}{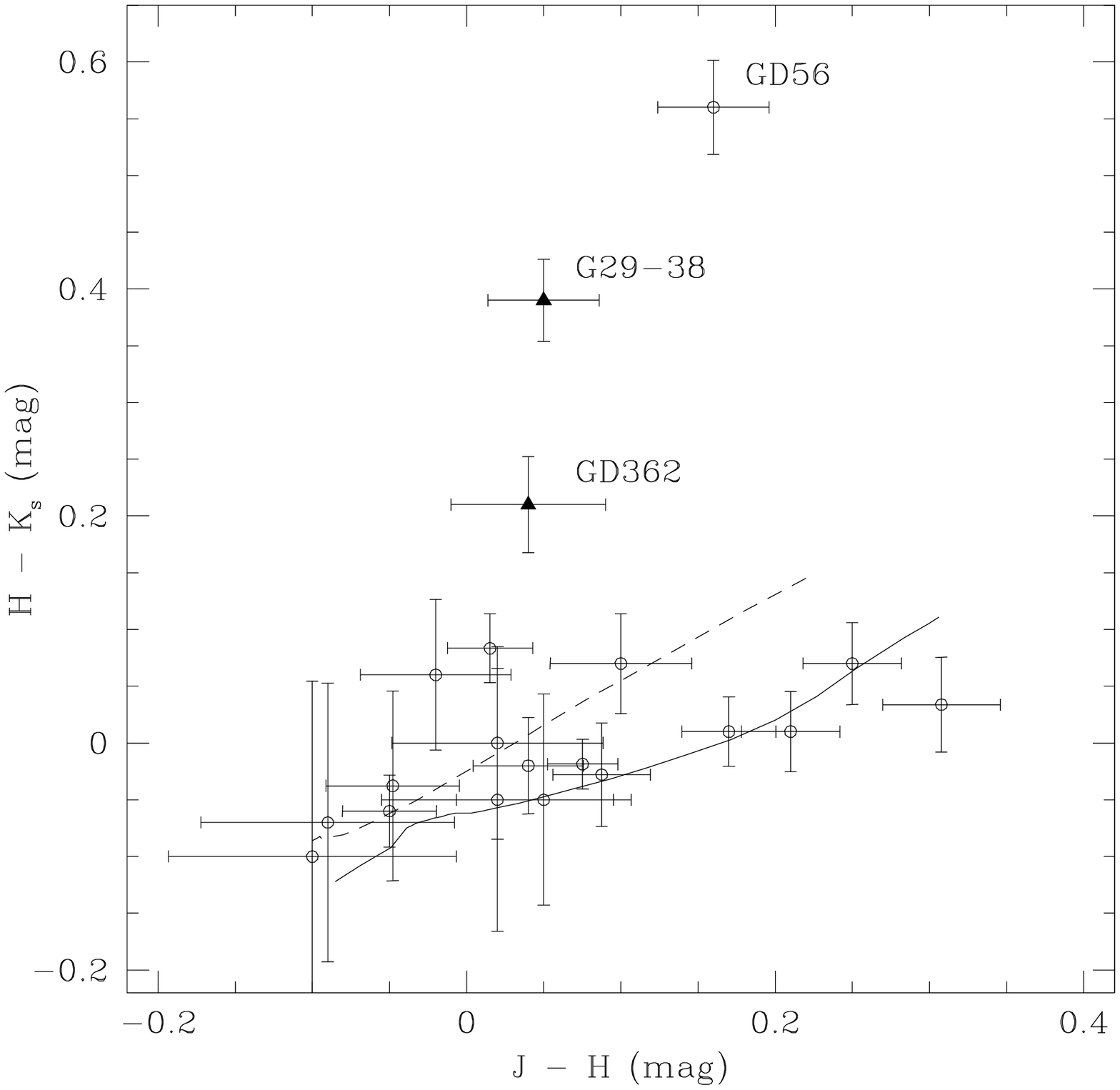}
\caption{
$J-H$ and $H-K$ color versus temperature and $J-H$ versus $H-K$ color-color diagrams for the observed DAZ stars.
The predicted sequences for DA (solid line) and DB (dashed line) white dwarfs are also shown. The
previously known white dwarfs with circumstellar debris disks (G29-38 and GD362) and the newly discovered white
dwarf with significant infrared excess (GD56) are labeled.}
\end{figure}

\clearpage
\epsscale{1}
\begin{figure}
\plotone{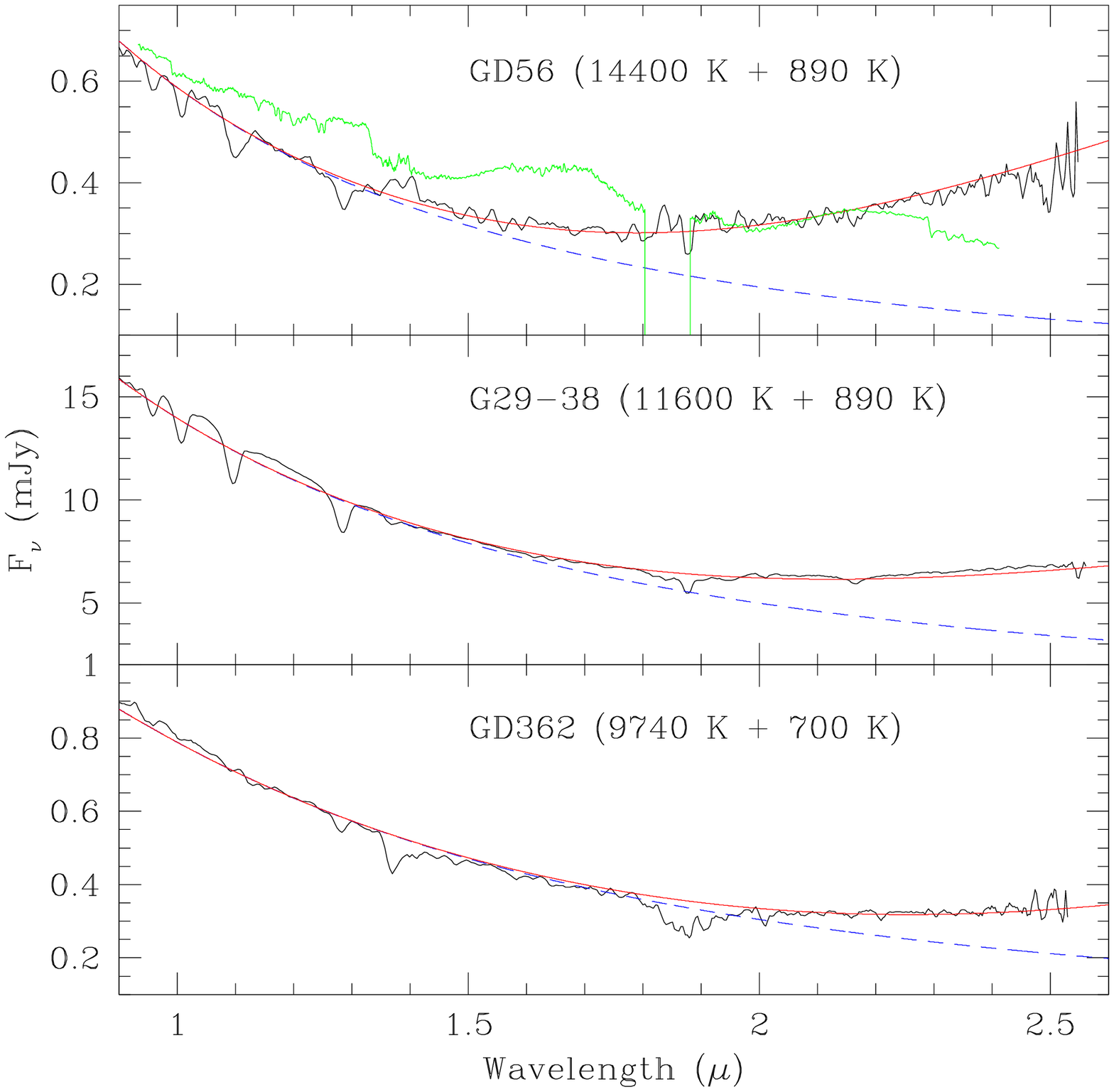}
\caption{The IRTF spectra of the white dwarfs with circumstellar debris disks (black lines). The expected
near-infrared fluxes from each star (dashed line) and composite white dwarf + dust templates (red line, both blackbodies) are shown
in each panel. The top panel also shows a composite blackbody + L3 dwarf template for GD56 (green line).} 
\end{figure}

\begin{figure}
\plotone{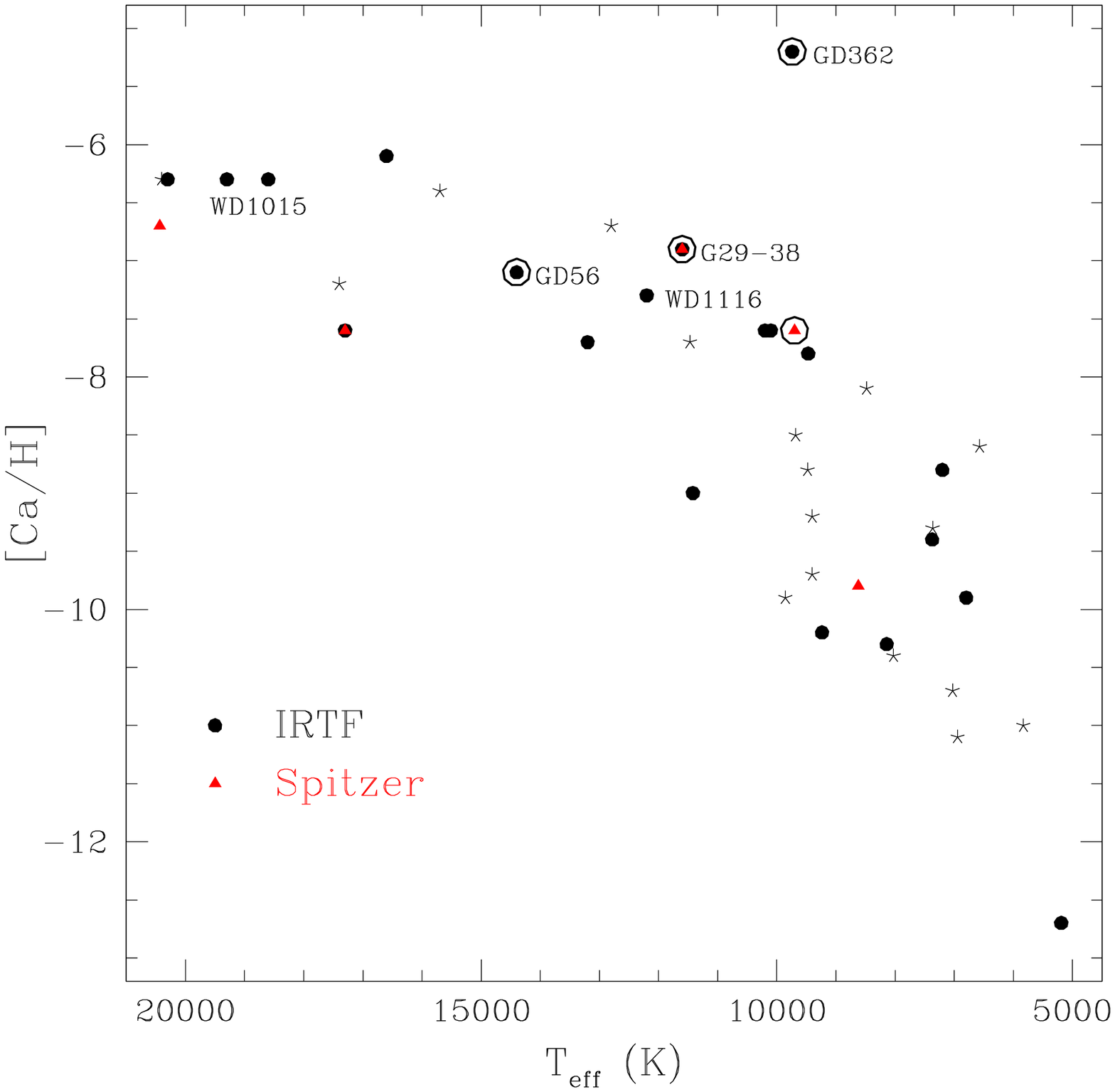}
\caption{Calcium abundances versus effective temperatures for the objects observed at the IRTF (filled circles) and
Spitzer/IRAC (filled triangles; from Reach et al. 2005 and von Hippel et al. 2006). The rest of the DAZs
from Berger et al. (2005) are shown with star symbols. White dwarfs with circumstellar debris disks are
marked with open circles.}
\end{figure}

\end{document}